\documentclass[aps,pra,floatfix,onecolumn]{revtex4}

\usepackage{bm}			    
\usepackage{amsmath}
\usepackage{amssymb}
\usepackage{latexsym}
\usepackage{amsfonts}
\usepackage{graphicx}
\usepackage{color}

\usepackage[textsize=notestyle]{todonotes}

\DeclareMathOperator{\Tr}{Tr}

\newcommand{\beq}{\begin{equation}}
\newcommand{\eeq}{\end{equation}}

\newcommand{\be}{\begin{equation}}
\newcommand{\ee}{\end{equation}}
\newcommand{\bc}{\begin{center}}
\newcommand{\ec}{\end{center}}
\newcommand{\esp}[1]{\left< #1 \right>}
\newcommand{\cor}[1]{\left[ #1 \right]}
\newcommand{\pare}[1]{\left( #1 \right)}
\newcommand{\key}[1]{\left\{ #1 \right\}}
\newcommand{\ben}{\begin{eqnarray}}
\newcommand{\een}{\end{eqnarray}}

\begin{document}

\title{Quantum transport in networks and photosynthetic complexes at the steady state}

\author{Daniel Manzano}
\email{daniel.manzano@uibk.ac.at}
\affiliation{Instituto Carlos I de Fisica Teorica y Computacional,
University of Granada, Av. Fuentenueva s/n, 18006, Granada, Spain}
\affiliation{Institute for Theoretical Physics,
University of Innsbruck,
Technikerstr.~25, A-6020 Innsbruck, Austria}
\affiliation{Institute for Quantum Optics and Quantum Information,
Austrian Academy of Sciences,
Technikerstr.~21A, A-6020 Innsbruck, Austria}

\date{Last update: \today}

\begin{abstract}
Recently, several works have analysed the efficiency of photosynthetic complexes
in a transient scenario and how that efficiency is affected by environmental noise.
Here, following a quantum master equation approach, we study the energy and
excitation transport in fully connected networks both in general and
in the particular case of the Fenna--Matthew--Olson complex.
The analysis is carried out for the steady state of the system where the excitation
energy is constantly ``flowing'' through the system. Steady state transport scenarios
are particularly relevant if the evolution of the quantum system is not conditioned
on the arrival of individual excitations.
By adding  dephasing  to the system, we analyse the possibility of noise-enhancement
of the quantum transport. 
\end{abstract}

\maketitle

\section{Introduction}
In the last years, quantum transport in photosynthetic complexes has become
an interesting field of study and debate.  An important part of this research 
focusses on the excitation transfer from the antennae that harvest the sunlight
to the reaction centre (RC) where the photosynthetic process takes place.
More concretely, for the  Fenna--Matthew--Olson (FMO) complex of green sulfur
bacteria, empirical evidence suggests that such transport is coherent
even at room temperature \cite{engel:nature07,collini:n10,panitchayangkoon:pnas10}.
These experiments show that the transient behaviour takes place on time scales
much shorter than the decoherence time due to the environment. Thus,
most of the recent analysis has focussed on the single-excitation
scenario in the transient regime obtained after pulsed photoexcitation
\cite{plenio:njp08,mohseni:jcp08,olayacastro:prb08,caruso:jcp09,rebentrost:njp09,
chin:njp10,wu:njp10,scholak:pre11,scholak:jpb11}.

Actually, there is a vivid debate about  the validity of the single-excitation picture 
for modelling the photosynthetic process {\it in vivo}. Photosynthesis in nature is a continuous process of absorption of energy
from a radiation field. As there is no specific measurement mechanism
that determines when the quanta of energy are effectively absorbed, some 
authors have argued that the photosynthetic complex and the radiation field
should evolve to a steady state
where the energy is constantly flowing through the system \cite{mancal:njp10,brumer:arxiv11}.
Some of the conclusions of \cite{brumer:arxiv11}, regarding the importance of a steady state picture, are summarized
 in the following paragraph:
 `The classical picture of the photon as a particle incident on the molecule, repeatedly initiating dynamics, also assumes a known photon arrival time. This too is incorrect and inconsistent with the quantum analysis insofar as no specific arrival time can be presumed unless the experiment itself is designed to measure such times'.
Also, it has been shown that some conclusions regarding the presence of entanglement in 
this kind of system rely on the assumption that the system is excited by a  single excitation Fock state. 
This state cannot be obtained just by weak illumination, and changing this assumption for 
a more realistic one changes dramatically the conclusions  \cite{tiersch:ptrsa12}.
These arguments makes it reasonable to analyse the natural photosynthetic processes also in 
other regimes, such as a steady state scenario. 

Moreover, quantum transport in a non-equilibrium steady state is an active field
in condensed matter physics. For ordered systems composed of qubits or harmonic oscillators,
it has been shown that it is possible to violate Fourier's law and thus achieve an infinite
thermal conductivity in the absence of noise \cite{manzano:arxiv11,asadian:arxiv12}.
This ballistic transport turns into a diffusive one, with finite conductivity,  
if noise is added to the system as a dephasing channel, reducing therefore the energy transfer.
That fact highlights the importance of the interaction with a dephasing 
environment for the energy transfer. The analysis of quantum transport can contribute to the design of artificial 
light-harvesting systems that are more efficient and robust \cite{blankenship:science11}.

Recently,  quantum transport in photosynthetic complexes 
has been analysed through different models  with different measures
of the efficiency, principally in the single-excitation regime.
In \cite{wu:njp10}, the dynamics of the FMO complex was analysed
by the use of a Markovian Redfield equation and by a generalized
Bloch--Redfield equation \cite{cao:jpc97}. The measure of efficiency
that they use is the average time that a single excitation spends
in the network before being absorbed by the sink.
The results show that the Redfield approach correctly describes the dynamics
of the system, but also that it fails to determine the optimal
dephasing ratio that minimizes the trapping time.
Moreover, this approach gives the unphysical results of a zero trapping
time in the limit of strong dephasing $\gamma\to \infty$. 
An analogous model was considered in \cite{chin:njp10}, with the difference
that the efficiency was quantified by the population of the sink in the long time limit.
Finally, Scholak {\it et al.} \cite{scholak:pre11,scholak:jpb11} have studied this problem in the
absence of a sink, in such a way that the only incoherent dynamics was
due to the presence of a dephasing environment. 
Here, the index for the quantification of the efficiency was the highest probability of finding the excitation
in the outgoing qubit in a time interval $[0,\tau]$, with $\tau$ being
related to the estimate of the duration of the excitation
transfer in real systems.
The authors conclude that the addition of noise can increase the efficiency,
but mainly in configurations that initially performed poorly.
Despite their differences, all these papers coincide in  analysing only the transient
behaviour, and not the steady state, and they use very different indexes for 
quantifying the efficiency of the system.

In this paper, we analyse the energy transfer in quantum networks and, specifically, in the FMO complex
in a steady state.
We show that the excitations move coherently through the system also in this regime.
The addition of a dephasing environment reduces, but does not destroy, the coherent
transport. We also analyse the change in efficiency due to such an environment.
The model we consider here is based on a quantum network connected to a thermal bath,
to model the absorption of energy from the radiation field, and to a sink,
that delivers the energy quanta to the reaction centre.
As a particular case, we analyse the FMO complex and similar fully connected networks. 
In this scenario, the system evolves to a non-equilibrium steady state,
where all the observables remain constant.  A similar framework has already been used 
to analyse entanglement in light-harvesting complexes in the transient regime \cite{caruso:pra10}.

As has been discussed before, several indexes of the efficiency are usually applied in order 
to calculate the efficiency of these kinds of system. Also, 
for the complete photosynthetic procces itself,  there is an important difference between analysing 
it by the use of quantum efficiency, that is, the 
average number of absorbed photons that finally give rise to photosynthetic products, and the energy efficiency. The second 
one is considered a more appropriate measure for comparing the efficiency of photosynthetic complexes with 
artificial light harvesting systems and for analysing the global procedure  \cite{blankenship:science11}. 
Because of that, we will use the energy transfer per unit of time, that is, the power, as our principal index of the 
efficiency of the systems. This measure will be compared with the excitation transfer, which corresponds to the quantum efficiency.  
We will show that, in general, they behave in a very different way, especially under the effect of noise.

The present paper is organized as follows.  In the next section, we introduce the details of the model
and of the master equation which describes its dynamics. In Section III, we introduce two indexes
for evaluating the efficiency, and we perform an analytical comparison between the two of them. 
Uniform and general networks are analysed in Section IV, while in Section V
we focus our attention on the FMO complex and related Hamiltonians.
Finally, in Section VI some conclusions are drawn.

\section{Description of the model and the master equation}

The energy transfer in photosynthetic complexes, such as the FMO complex, 
can be described by exciton dynamics. Such systems can be modelled 
as fully connected networks of two-level systems (qubits), and 
several recent works have  analysed photosynthetic processes by the use of this
framework. Most of these works have used a single excitation framework, 
different from the one used here. 
Since the FMO complex is composed of seven chromophores,
it should be modelled by a network of seven qubits.
To describe the absorption of energy from the antennae and the decay
to the reaction centre (RC), we use a Markovian quantum master equation
in Lindblad form \cite{breuer_02}.
The validity of this master equation has been numerically verified in
systems composed of harmonic oscillators \cite{rivas:njp10}, showing
that it is accurate for small coupling even in the low and high temperature regimes.
In the transient regime, similar master equations have been previously
used to describe the dynamics of the FMO complex and to analyse 
the effects of noise on the quantum dynamics \cite{chin:njp10,wu:njp10}.

The quantum evolution of the network is determined by a Hamiltonian of the form 

\be
H=\sum_{i=1}^7 \hbar \omega_i \sigma_i^+ \sigma_i^- + \sum_{\substack{i,j=1 \\  j\ne i}}^7 \hbar g_{ij} \left( \sigma_i^+ \sigma_j^- +  \sigma_i^- \sigma_j^+  \right), 
\ee
where $\sigma_i^\pm$ are the raising and lowering operators that act on qubit $i$,
$\hbar \omega_i$ are the one-site energies, and $g_{ij}$ 
represents the coupling between qubits $i$ and $j$. 

The interaction of the photosynthetic complex with the environment can be divided into four different processes, 
and each of them is modelled by a different Lindblad superoperator. According to the empirical results from \cite{adolphs:bj06},  
the absorption of energy from the antennae populates principally 
site $1$, with a non-negligible population of site $6$. For simplicity, this process is modelled in our paper by a thermal bath 
connected to site $1$. The delivery of the excitation energy from the complex to the RC is mediated by site $3$. As this is a irreversible 
process, we model it by a sink, a zero temperature thermal bath, which removes the excitations from this site in an 
incoherent way. Also, photosynthetic complexes 
are not isolated from the surrounding environment.  They interact with other biological components, an interaction that 
 {\it in vivo} happens at room temperature. That leads to two different effects on the system. First, the loss of coherence in 
 the transport due to the dephasing induced in the system, and second, the absorption of excitations by the environment.

The injection of excitations by the thermal bath acting on the first qubit is modelled  by the Linblad superoperator

\be \label{eq:lindblad}
\mathcal{L}_{\text{th}} \rho
=
\Gamma_{\text{th}} (n+1) \left( \sigma^-_1 \rho \sigma^+_1 -\frac{1}{2} \left\{\sigma^+_1\sigma^-_1,\rho\right\} \right) 
+\Gamma_{\text{th}} n \left( \sigma^+_1 \rho \sigma^-_1 -\frac{1}{2} \left\{\sigma^-_1\sigma^+_1,\rho\right\} \right),
\ee
where the parameter $\Gamma_{\text{th}}$ represents the strength of the coupling between the quantum system and the environment.
 As there are no empirical estimates of this parameter, we take $\Gamma_{\text{th}}=1\; \text{cm}^{-1}$ 
through the paper. The parameter $n$ is the mean number of excitations with frequency $\omega$  in the bath.

The delivery of energy from the system to the RC is modelled by a second Lindblad
superoperator, which models a zero temperature thermal bath. This bath is usually referred to as a sink. 

\be
\mathcal{L}_{\text{sink}} \rho = \Gamma_{\text{sink}} \left( \sigma^-_3 \rho \sigma^+_3 -\frac{1}{2} \left\{\sigma^+_3\sigma^-_3,\rho\right\} \right).
\ee

This term describes the irreversible decay of the excitations to the reaction centre.
When the excitation is absorbed by the sink, it triggers a charge separation event and can not go back to the complex.
We assume that this process is faster than the system dynamics and, because of that,
the sink does not saturate. So, it can be described by a Markovian approach.
Again, the coupling strength  $\Gamma_{\text{sink}}$ has not been estimated
from experimental data, and we choose $\Gamma_{\text{sink}}=1 \text{ cm} ^{-1}$.

The interaction between the complex and its surrounding environment has two different effects. First, it reduces the quantum coherence of the 
system. This is modelled in our master equation by the term 

\be
\mathcal{L}_\text{deph} \rho=\gamma \sum_{i=1}^7 \pare{\sigma_i^+\sigma_i^-\rho\sigma_i^+\sigma_i^- - \frac{1}{2}\key{\sigma_i^+\sigma_i^-,\rho}}.
\ee

This interaction does not change the mean number of excitations in the 
system but, as we will see in next section, it can affect the its mean energy, and so the energy flux. Also, 
this is the interaction that can improve the efficiency of the system by removing the destructive coherences that delay 
the transmission of the excitation to the third site. The parameter $\gamma$ represents the strength of the interaction 
between the complex and the dephasing environment. This will be the free parameter we use in this paper in order to 
optimize the efficiency of the system. 

Finally, the system is also susceptible to  a radiative decay process that transfers the excitations from 
the complex to the environment. This process effectively reduces the mean number of excitations in the 
system together with the mean energy. 

\be
\mathcal{L}_{\text{diss}} \rho = \Gamma_{\text{diss}}\sum_{i=1}^7 \left( \sigma^-_i \rho \sigma^+_i -\frac{1}{2} \left\{\sigma^+_i\sigma^-_i,\rho\right\} \right).
\ee

Again, the coupling parameter $ \Gamma_{\text{diss}}$ has not been inferred from experimental data, so we will check 
different values of it in order to analyse the noise-enhancement under different kinds of dissipative environments. 

The complete time evolution of the density matrix of the system is described by the master equation 

\be\label{eq:me}
\dot{\rho}= -\frac{i}{\hbar} \cor{H,\rho} + \mathcal{L}_{\text{th}} \rho+ \mathcal{L}_{\text{sink}} \rho+\mathcal{L}_\text{deph} \rho+
\mathcal{L}_{\text{diss}} \rho .
\ee

The steady state occurs in the long time limit, and satisfies the condition $\dot{\rho}=0$, meaning that the density
matrix is stationary.

Our analysis has been performed in  two different regimes, depending on the rate of  excitation injection.
First, we study the low-temperature case, by choosing  $n=1$. This choice corresponds to 
a slow injection of excitations into the system, as should be the case for the FMO
complex under weak illumination.
In this case, the injection of energy is so slow that the probability of finding 
more than one excitation at the same time in the network is almost negligible (but not excluded).
Second, we simulate a high temperature environment, $n=100$, where there is
a higher probability of finding more than one excitation inside the system at the same time.
As we will see in the following sections, these different situations lead to different results.

\section{Energy and excitation fluxes}

In order to evaluate the efficiency of these systems,
we consider two different indexes.
First, we observe that the net energy transfer across the system is quantified
by the time derivative of the expectation value of the Hamiltonian, 

\be
\dot{E}=\frac{d}{dt}\esp{H}= \Tr \left( H \dot\rho \right).
\ee

By using the master equation (\ref{eq:me}) and the fact that, in the steady state,
the mean energy of the system is conserved, we obtain an expression
for the energy exchanged through the different environmental channels.

\be\label{eq:heat}
0=\Tr \pare{H \mathcal{L}_{\text{th}} \rho} + \Tr \pare{H \mathcal{L}_{\text{sink}} \rho} +\Tr \pare{H \mathcal{L}_\text{deph} \rho}
+ \Tr \pare{H \mathcal{L}_\text{diss} \rho} =: J_{\text{th}}+J_{\text{sink}}+ J_{\text{deph}}+J_{\text{diss}}, 
\ee
where $J_{\text{th}}$ represents the energy flux from the thermal environment to the system,
$J_{\text{sink}}$ from the system to the sink,  $J_{\text{deph}}$ between
the system and the dephasing environment, and $J_{\text{diss}}$ is the energy loss due 
to the decay of the excitations. 

As our main interest is to quantify the energy that flows from the system to the sink,
we use $J\equiv J_{\text{sink}}$ as our first index to measure the efficiency.
This expression has been applied in previous papers in order to analyse the efficiency of
 quantum refrigerators 
\cite{linden:prl10} and to study Fourier's law in quantum systems
\cite{manzano:arxiv11,asadian:arxiv12}.

The second measure of the efficiency that we use is the excitation flux, defined
as the number of excitations incoherently absorbed by the sink per unit time.
For a time interval $[0,\tau]$, it is given by

\be
p_{\text{sink}}=\Gamma_{\text{sink}} \int_{0}^{\tau} P dt,
\ee
where $P\equiv \esp{\sigma_3^+ \sigma_3^-}$ is the population of the
third site, obviously time-independent in the steady state.
This fact allows us to consider the population of the third 
site as a measure of the speed of the excitation transfer to the RC.
These two measures, the energy transfer and the population of the third site,
are not equivalent since they measure slightly different quantities.
Indeed, the results about the efficiency of the systems are different
depending on the measure that is used.

In order to relate these two quantities, it is is useful to decompose the Hamiltonian:

\be
H=H_3 + \sum_{\substack{j=1 \\  j\ne 3}}^7 H_{3j} + \tilde{H},
\ee 
where $H_3=\hbar \omega_3 \sigma_3^+ \sigma_3^-$,  $H_{3j}=\hbar g_{3j} \pare{\sigma_3^+\sigma_j^- + \sigma_3^-\sigma_j^+}$, 
and $\tilde{H}$ represents the part of the Hamiltonian that does not involve qubit three. Hence, $J$ can be expressed by 

\be
J= \Tr (H_3 \mathcal{L}_{\text{sink}})+ \sum_{\substack{j=1 \\  j\ne 3}}^7  \Tr (H_{3j} \mathcal{L}_{\text{sink}})
+ \Tr (\tilde{H} \mathcal{L}_{\text{sink}}),
\ee
and the last term is null due to the fact that $\Tr (\mathcal{L}_{\text{sink}})=0$. A straightforward evaluation of this expression 
allows of expressing the energy transfer to the sink as a function of the population of the third site and the coherences between 
this qubit and all the others. 

\be\label{eq:flux}
J=-\Gamma_{\text{sink}}\hbar \pare{  - \omega_3 P -\sum_{\substack{i=1 \\  i\ne 3}}^7 
\frac{g_{i3}}{2} \pare{\esp{\sigma_3^+ \sigma_i^-} + \esp{\sigma_i^- \sigma_3^+}}  }.
\ee

As $\esp{\sigma_3^+ \sigma_i^-}=\esp{ \sigma_3^- \sigma_i^+ }^*$, the heat transfer will depend  only on the real part of the 
next-neighbours coherences. 
It has been proved that in a linear chain, with equal one-site energies and 
couplings, these coherences are purely imaginary and the heat flux depends only on the population \cite{manzano:arxiv11}. 
In the general case, these coherences will be nonzero and they can contribute in a positive or negative way to the energy flux. 
It is clear from Eq. (\ref{eq:flux}) that there is a strong connection between the energy and the population fluxes. It is also clear 
that these measures are related but not equivalent. 

In a similar way, the expression of the energy flux due to the dephasing environment can be calculated 

\be
J_{\text{deph}}=  \sum_{\substack{i,j=1 \\  j>i}}^7 - \frac{\gamma \hbar g_{ij}}{2} \pare{ \esp{\sigma_i^- \sigma_j^+} +\esp{\sigma_i^+ \sigma_j^-} }.
\ee

Again, in the concrete case of a uniform chain, these next-neighbours coherences are purely imaginary and because of that this term vanishes. 
In a general fully connected network, these elements are in general complex and there is an interchange of energy between the network and the environment 
due to the dephasing channel. That effect happens because the environment projects the system onto a basis that is not composed of
  the Hamiltonian's eigenvectors. That means a reduction of the elements that are not eigenvectors of a single 
  site basis. As the eigenvalues of the Hamiltonian are usually composed of these non-local terms, this interaction can effectively reduce or increase
  the energy inside the system. Recently, the energy cost of quantum projective measurements has been analysed 
  and related to the work value of the acquired information \cite{jacobs:pre12}. 
 Measurements can change the energy of the network and, in the steady state, that will lead to an energy flux. In a similar 
way, the presence of dissipation in the system reduces both the excitation and energy fluxes.

\section{General networks}

First, we analysed the case of a general fully connected network where the excitations are injected by  a thermal reservoir and
delivered to a sink. For that, we have calculated the energy 
end excitation rates as functions of the dephasing ratio for a fully connected homogeneous network in both the low and high temperature regime. 
Both the one-site energies and the couplings between the qubits are equal and they have the value
 $\hbar \omega_i=\hbar g_{ij}=1 \text{ cm}^{-1}\; \forall \, i,j$. 
The couplings with the thermal environment and the sink are  $\Gamma_{\text{th}}=\Gamma_{\text{sink}}=1\; \text{cm}^{-1}$. 
The analysis is made for different values of the dissipation rate $\Gamma_{\text{diss}}$, in order to analyse the effects of 
a dissipative environment on the transfer. 
 
\begin{figure}[ht!]
\bc
\includegraphics[scale=0.4]{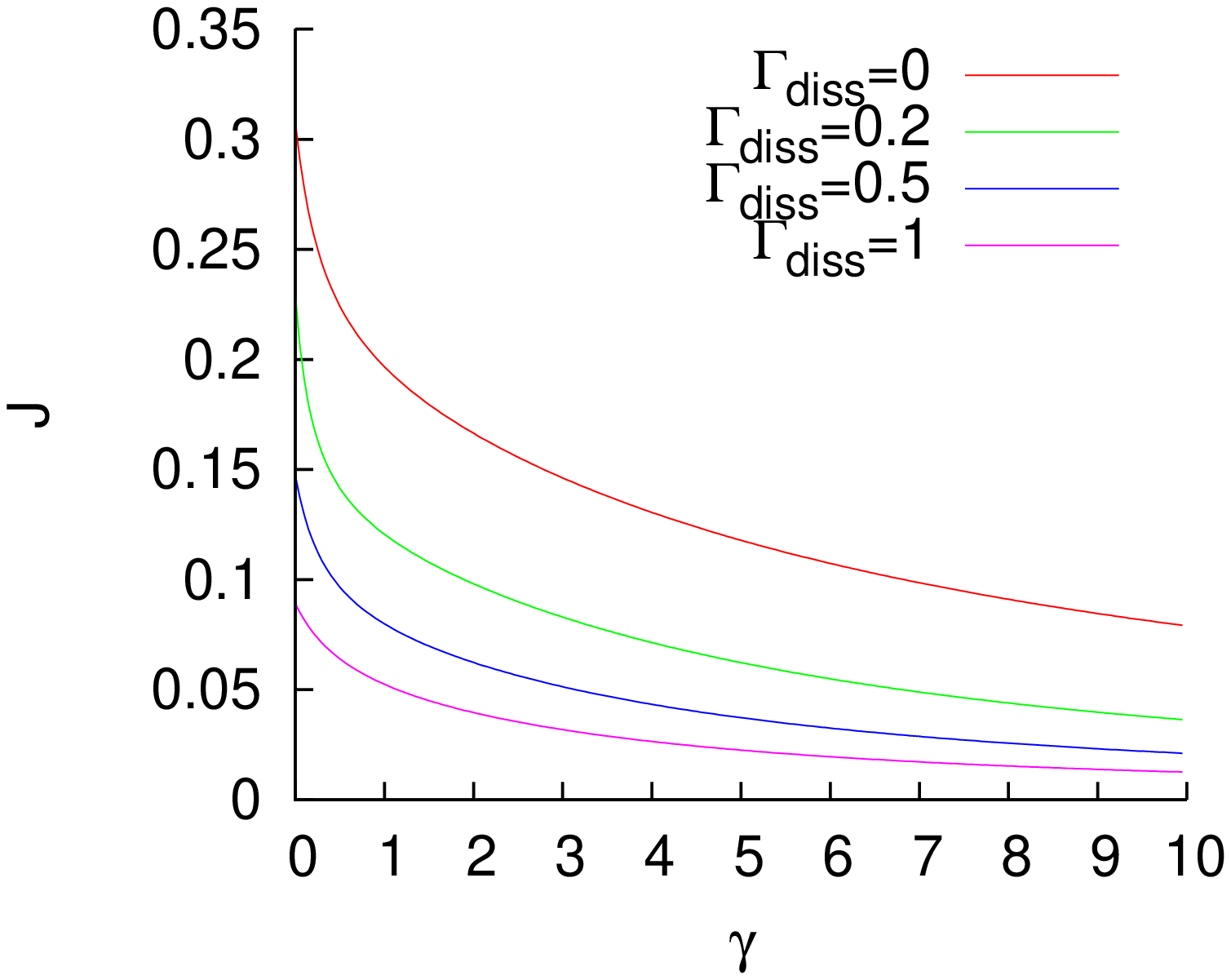} 
\includegraphics[scale=0.4]{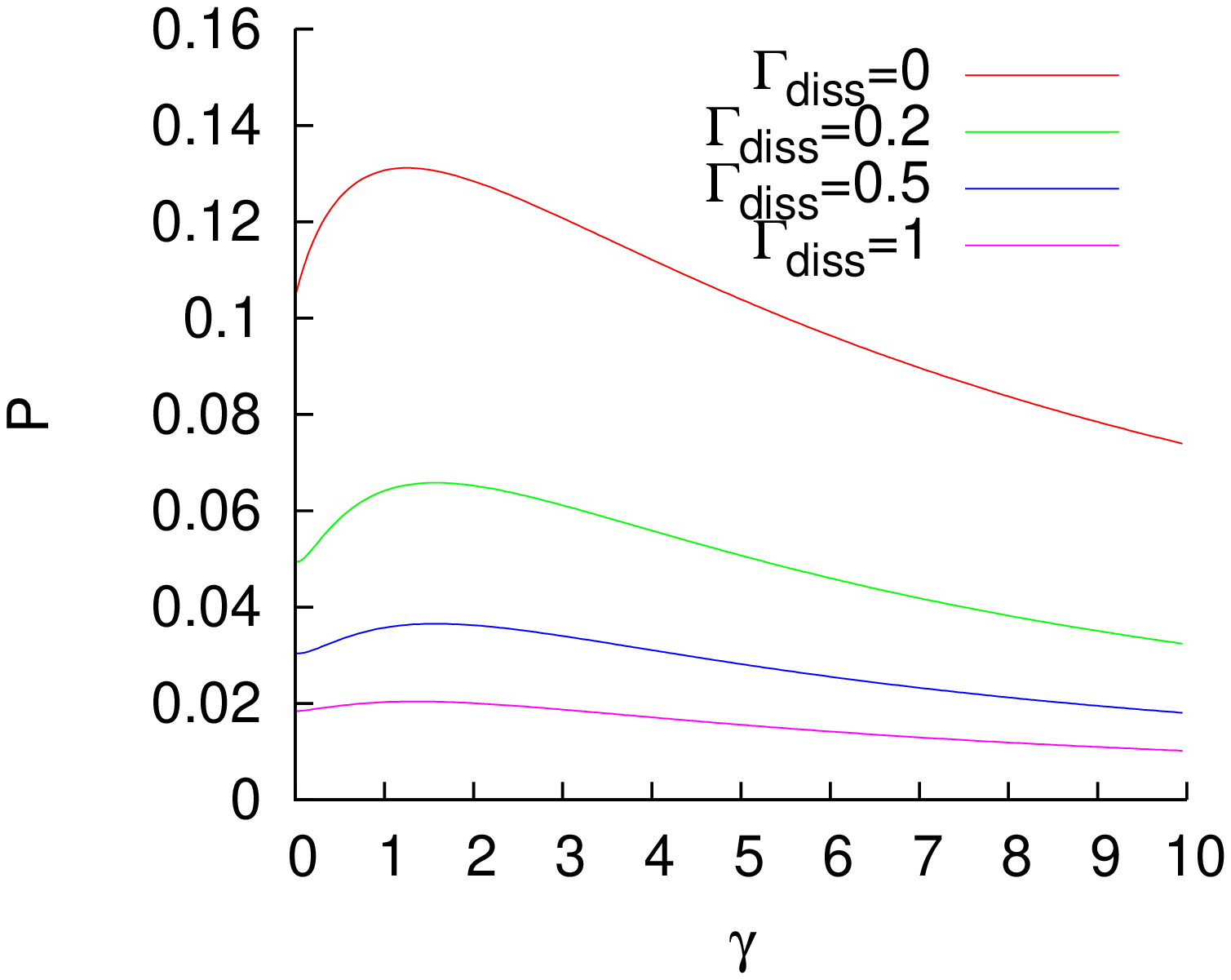} 
\ec
\caption{Excitation (left) and energy (right) fluxes for a homogeneous fully connected network, with  $\hbar \omega_i=\hbar g_{ij}=1 \text{ cm}^{-1}\; \forall \, i,j$, as functions of the dephasing ratio 
$\gamma$,  for a low-temperature thermal bath, $n=1$, and different dissipation rates $\Gamma_{\text{diss}}$. Units of cm$^{-1}$.}
\label{fig:newfig1}
\end{figure}

The results for a low temperature thermal bath ($n=1$), and several dissipation rates ($\Gamma_{\text{diss}}=0,0.2,0.5,1\;\text{cm}^{-1}$),
 are shown in Fig. \ref{fig:newfig1}. The two different measures of efficiency, the energy and excitation transfer, exhibit 
very different behaviours when the dephasing parameter increases. 
The excitation transfer increases for small values of the dephasing ratio and decreases for higher ones. 
It is clear that the excitation transfer at the steady state can be improved by the addition of external noise to the system. This effect is 
due to the reduction of destructive interferences that inhibit the transport of the excitation to the qubit coupled to the sink,
 as is explained in \cite{chin:njp10}. 
Similar results have been obtained 
 in the transient regime \cite{chin:njp10}. On the other hand, the energy transfer is always reduced if a dephasing environment 
 acts on the system. That means that, even 
where the noise can enhance the number of particles arriving at the sink per unit time, it can at the same time reduce the energy 
transferred to it. This difference  is due to the reduction of the coherences in Eq. (\ref{eq:flux}).
For high values of the dephasing rate $\gamma$, the transport is reduced, 
due to a quantum Zeno effect, that avoids the coherent transport.
 The optimal dephasing ratio that maximizes the excitation transfer is $\gamma_\text{uniform}^* =1.25 \pm 0.1 \text{ cm}^{-1}$. For 
 the energy transfer, the optimal rate is to have no dephasing at all. 
 
\begin{figure}[ht!]
\bc
\includegraphics[scale=0.4]{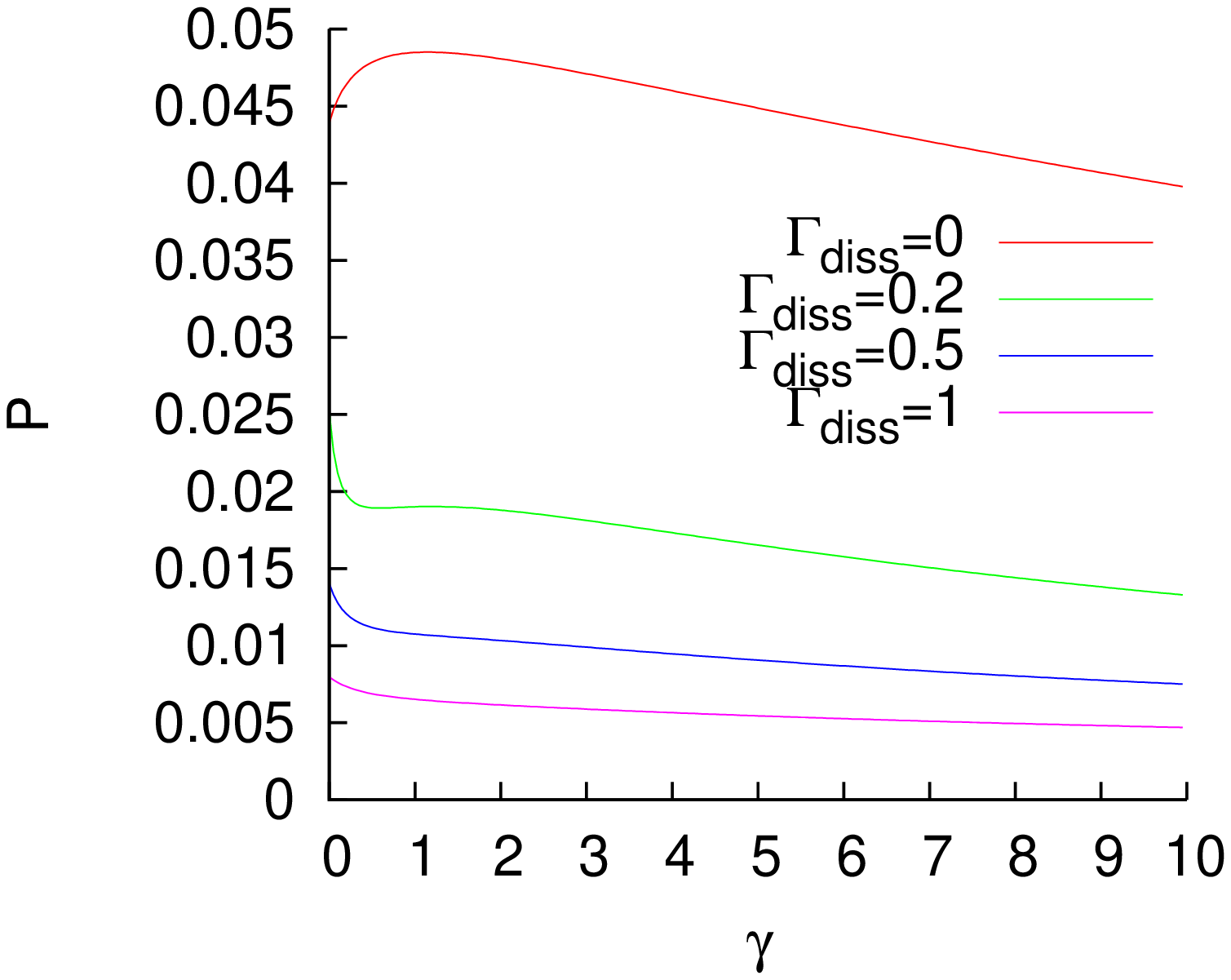} 
\includegraphics[scale=0.4]{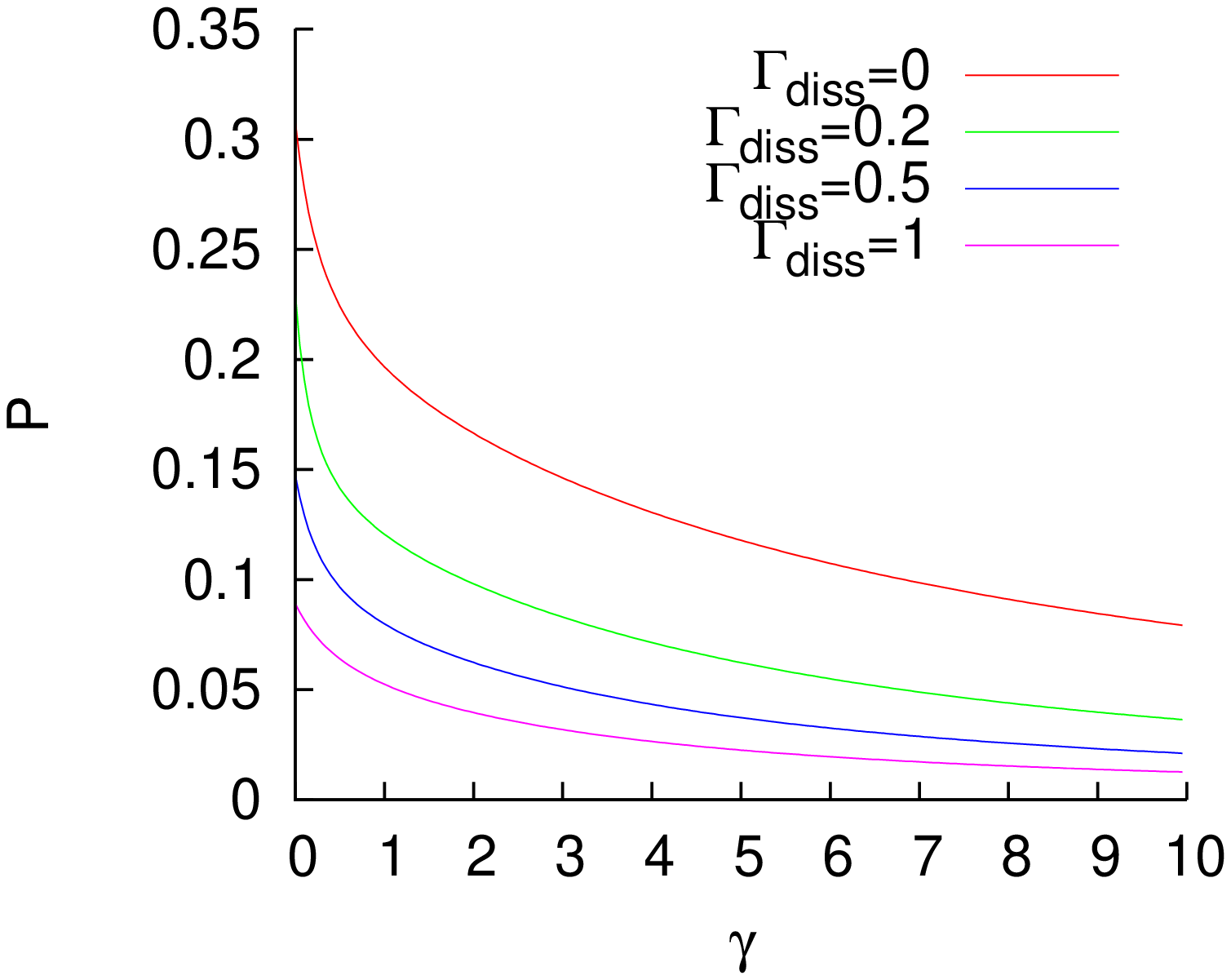} 
\ec
\caption{Excitation (left) and energy (right) fluxes for a homogeneous fully connected network,  $\hbar \omega_i=\hbar g_{ij}=1 \text{ cm}^{-1}\; \forall \, i,j$,  with a thermal 
mean excitation number $n=100$, and different dissipation rates $\Gamma_{\text{diss}}$. Units of cm$^{-1}$.}
\label{fig:newfig2}
\end{figure}

 For a high temperature bath ($n=100$), we have similar results, as is displayed in 
 Fig. \ref{fig:newfig2}. For this case, both fluxes are higher than in the low temperature regime, as is to be expected, but the effect of 
 dissipation is different. If the system is under the effect of a highly dissipative environment, as $\Gamma_{\text{diss}}=1\;\text{cm}^{-1}$, 
 the efficiency  of the excitation flux can not be improved by the addition of dephasing. 
 The optimal dephasing ratio is independent of the temperature of the bath for no dissipation, and in the low temperature regime 
 it is independent of the dissipation. That indicates that it could be a general property of the Hamiltonian. 
 For a highly dissipative bath, the  noise-enhancement of the excitation transfer progressively disappears and, because of that, 
 there is no optimal dephasing rate.

\begin{figure}[ht!]
\bc
\includegraphics[scale=0.4]{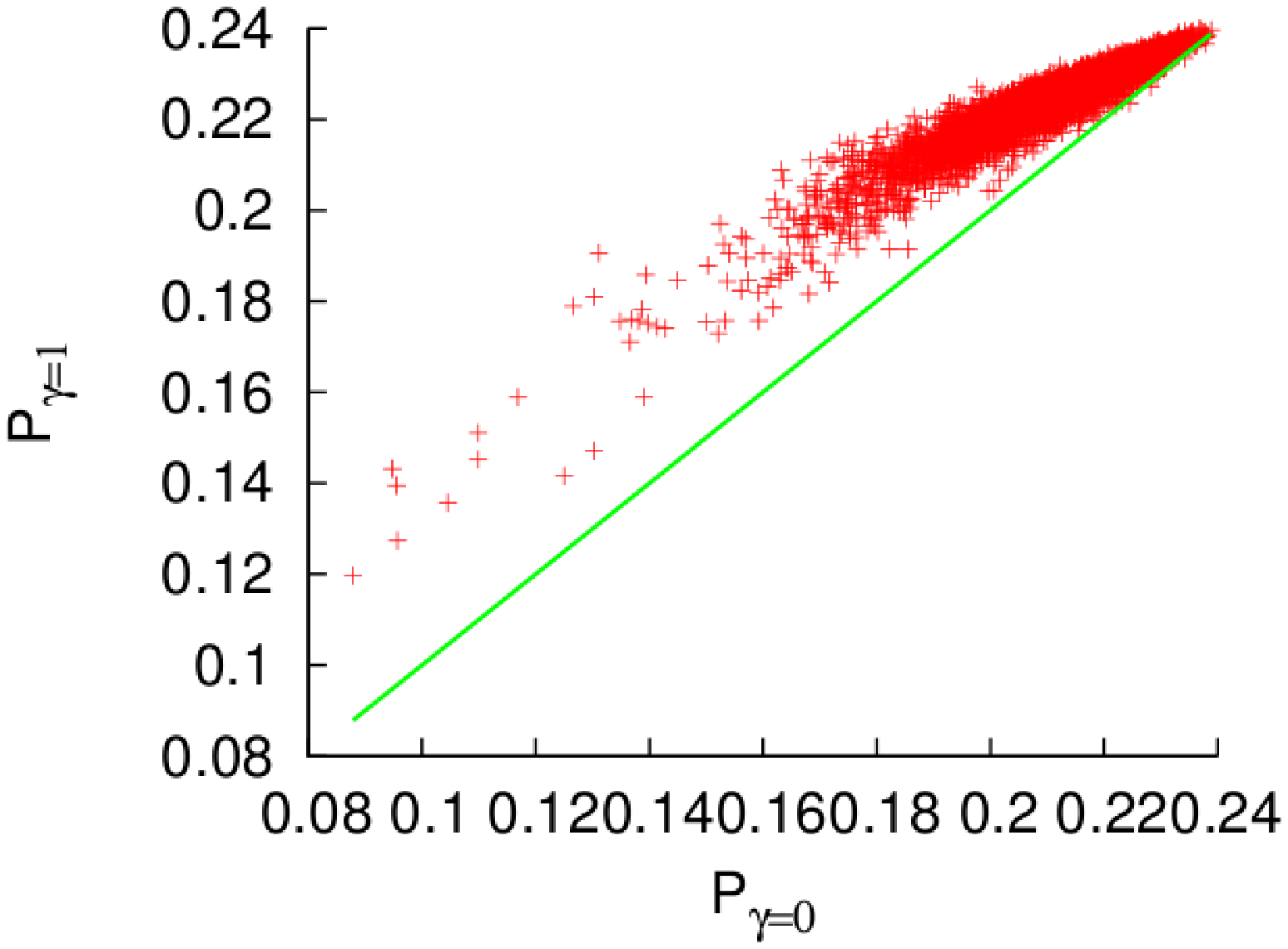} 
\includegraphics[scale=0.4]{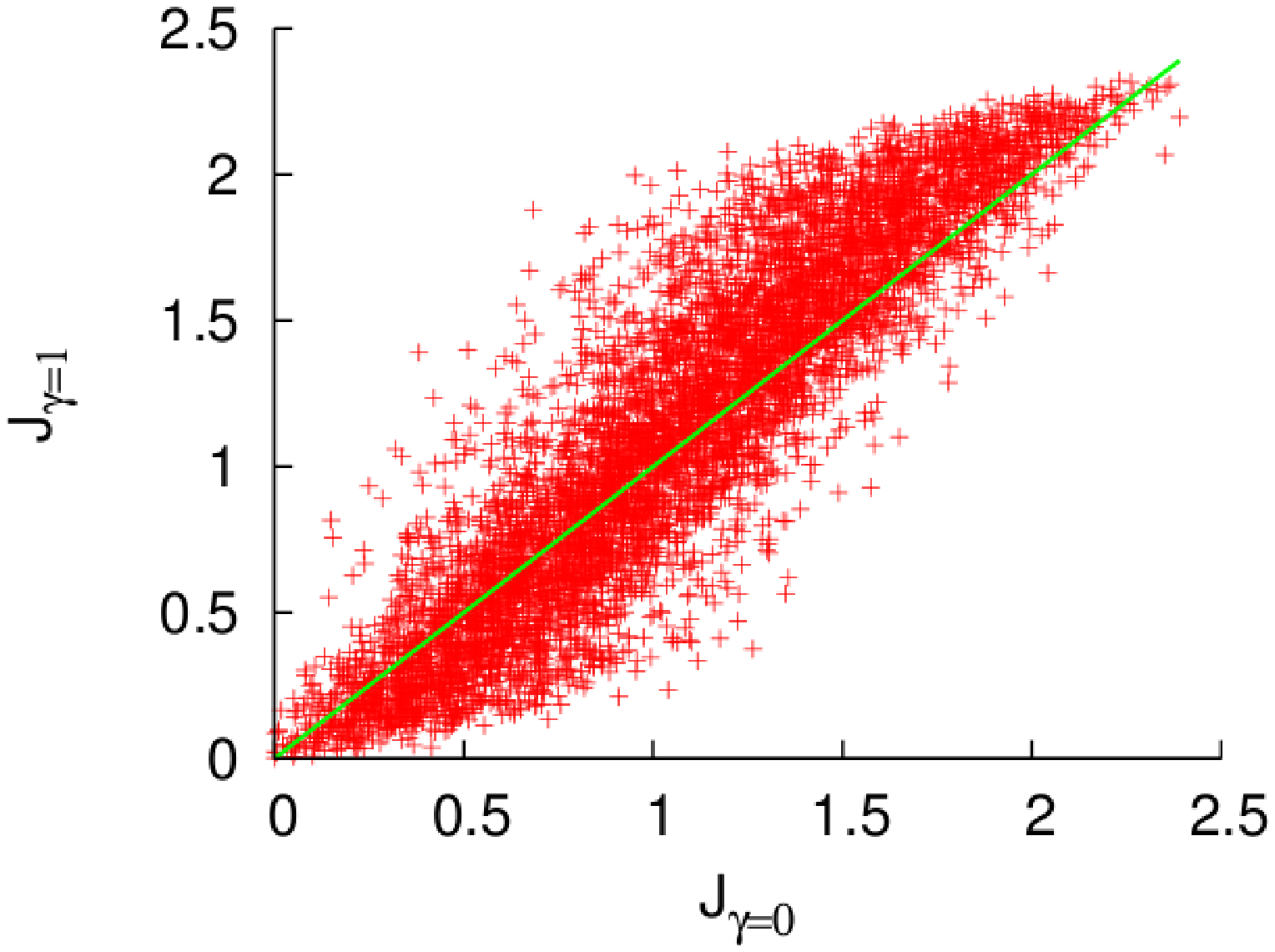} 
\ec
\caption{Energy (left) and excitation (right) fluxes for random Hamiltonians in the low-temperature regime ($n=1$), for a dephasing ratio 
$\gamma=1$ as a function of the fluxes without dephasing. The green line separates the configurations with enhancement  and depression 
of the transfer. Units of cm$^{-1}$. }
\label{fig:fig5}
\end{figure}

For analysing a more general case, we have also generated 7000 random Hamiltonians, by Monte Carlo simulation, where the one-site 
energies and the couplings between the qubits are randomly selected from a uniform 
distribution, with $\hbar \omega_i\in [0,10]$ and $\hbar g_{ij}\in[-10,10]$ cm$^{-1}$.
Again, the heat and the excitation transfer exhibit very different behaviours. In Fig. \ref{fig:fig5},   the heat and population 
fluxes are plotted for a small dephasing ratio, $\gamma=1$,  as a function of the fluxes for the same systems without dephasing, 
 in the low temperature regime.
 For the heat transfer, the addition of noise to the system can either increase or decrease the efficiency of the system. 
This improvement is smaller when the system is highly efficient. That shows that the most efficient configurations are also the most robust against 
the addition of noise, and they are very difficult to improve. The excitation transfer to the sink is improved for all the analysed Hamiltonians, 
again this improvement is less when the efficiency of the system is greater. 
These results are compatible with the conclusions of \cite{scholak:pre11}, 
where it is shown that in the single excitation picture, systems with low efficiency are more suitable for improvement by a dephasing 
channel than are the highly efficient ones.

\begin{figure}[ht!]
\bc
\includegraphics[scale=0.4]{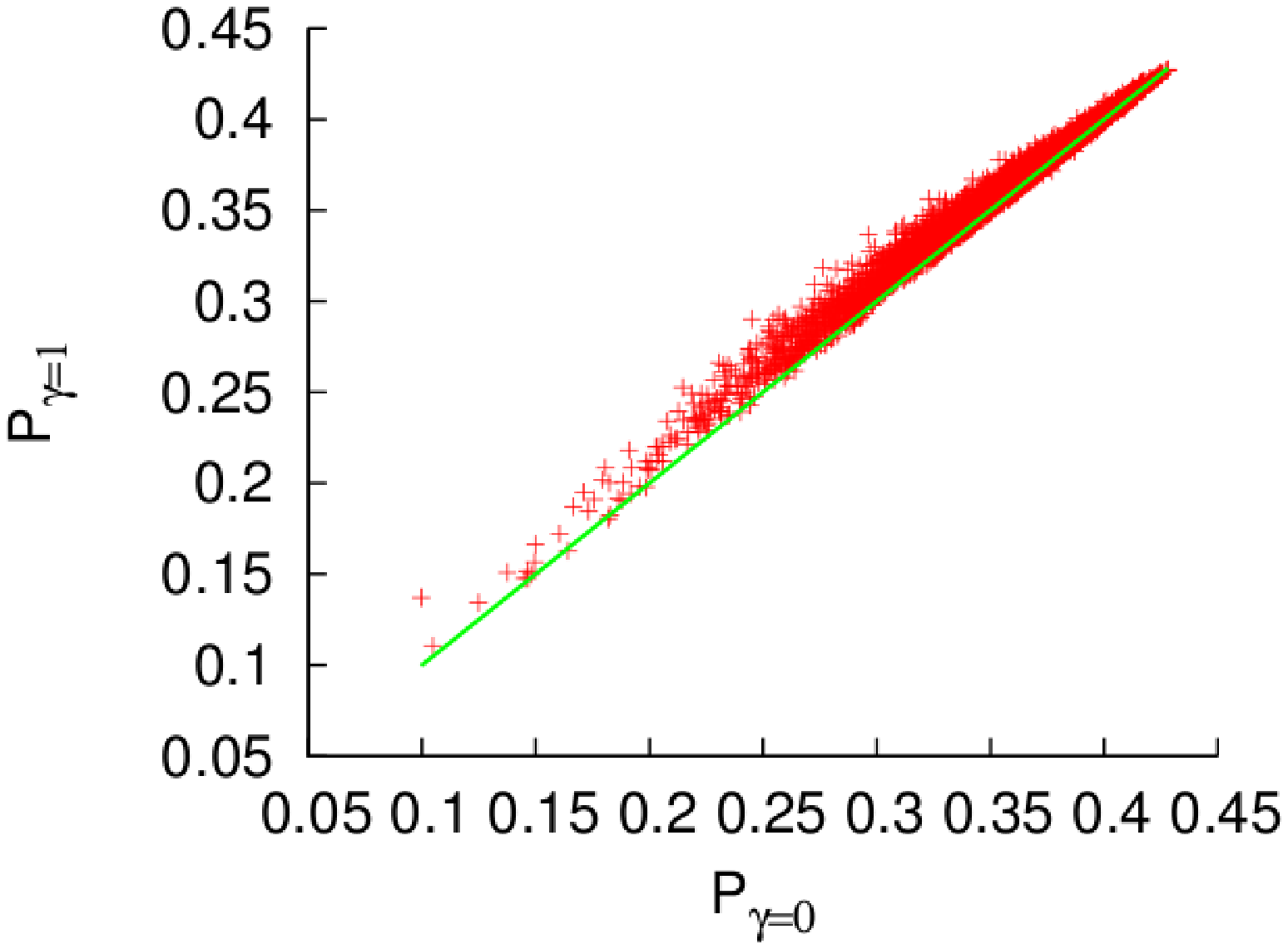}
\includegraphics[scale=0.4]{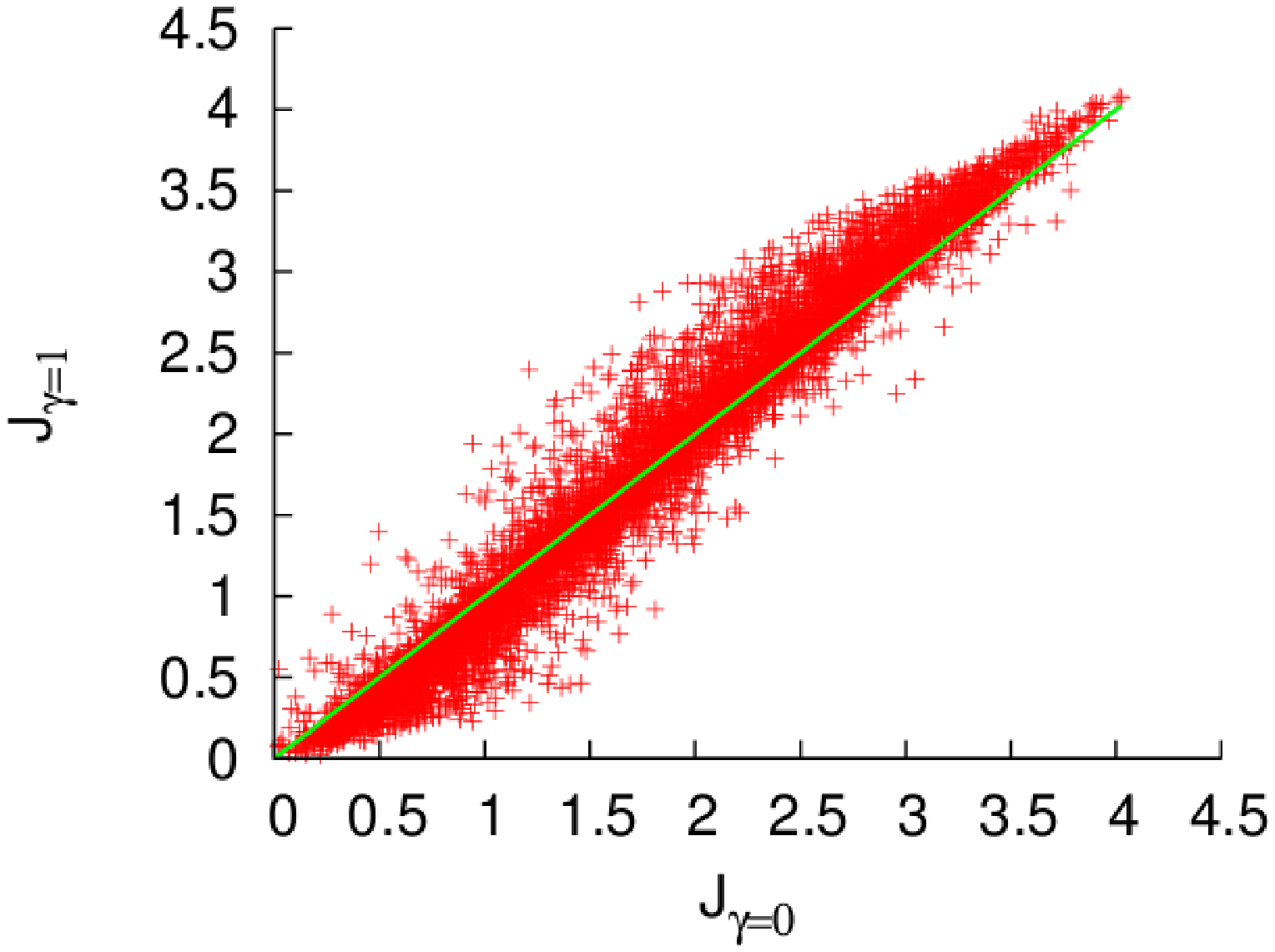} 
\ec
\caption{Energy (left) and excitation (right) fluxes for random uniformly distributed 
Hamiltonians in the low-temperature regime ($n=100$), for a dephasing ratio $\gamma=1$, as functions of the fluxes without dephasing. 
The green line separates the configurations with enhancement  and depression 
of the transfer. Units of cm$^{-1}$.}
\label{fig:fig6}
\end{figure}

The results in the case of a high energy thermal bath  are very similar, as is displayed in Fig. \ref{fig:fig6}. There, the enhancement is smaller 
than for $n=1$ and the population transfer can also be reduced, but only to  a small degree. That implies that the energy transfer is more stable 
under the effects of noise for systems with high number of excitations than for ones in which this number is smaller. Again, the most 
efficient configurations are more robust against the effect of noise. 

Both in the low and high temperature regimes, the effects of dissipation in these simulations are similar. The presence of dissipation 
reduces both the fluxes and the improvement possible with a dephasing noise. The results are very similar to the ones displayed in Figs. 
\ref{fig:fig5} and \ref{fig:fig6} and are not shown for simplicity.

The differences between the energy and heat transfer come from the fact that the energy transfer depends both on the population of the 
outgoing site and the coherences between it and the others qubits. Even if the dephasing channel increases the population of this 
qubit, increasing consequently the transfer of excitations to the sink, it also reduces the amount of coherence between 
qubits. These two effects compete  to determine 
whether the energy transfer will be improved or depressed. The relation between the one-site energies and the couplings play an essential and 
non-trivial role in this effect. 

\section{Photosynthetic complexes}

To analyse light-harvesting biological systems, we study the FMO protein complex in green sulfur bacteria. This complex 
is assumed to have seven chromophores and, because of that, it can be modelled as a network of seven sites.
 We use the experimental Hamiltonian given in \cite{adolphs:bj06}, tables 2 MEAD and 4 (trimer). In cm$^{-1}$, this Hamiltonian reads: 

\be
H_{\text{FMO}}=\left(
\begin{array}{ccccccc}
12445 & -104.0 & 5.1 & -4.3 & 4.7 & -15.1 & -7.8 \\
-104.0 & 12450 & 32.6 & 7.1 & 5.4 & 8.3 & 0.8 \\
5.1 & 32.6 & 12230 & -46.8 & 1.0 & -8.1 & 5.1 \\
-4.3 & 7.1 & -46.8 & 12355 & -70.7 & -14.7 & -61.5\\
4.7 & 5.4 & 1.0 & -70.7 & 12680 & 89.7 & -2.5 \\
-15.1 & 8.3 & -8.1 & -14.7 & 89.7 & 12560 & 32.7 \\
-7.8 & 0.8 & 5.1 & -61.5 & -2.5 & 32.7 & 12510
\end{array}
\right).
\ee
 
As for this Hamiltonian the one-site energies are two orders of magnitude higher than the couplings, we can expect that they should  play a  
more relevant role for the energy flux. In this concrete case, the energy and excitation transfer are very similar,  in contrast to the 
general networks analysed before. As there are no empirical measurements of the coupling between the complex and the antennae or the 
RC, we choose $\Gamma_{\text{th}}=\Gamma_{\text{sink}}=1\; \text{cm}^{-1}$.

\begin{figure}[ht!]
\bc
\includegraphics[scale=0.4]{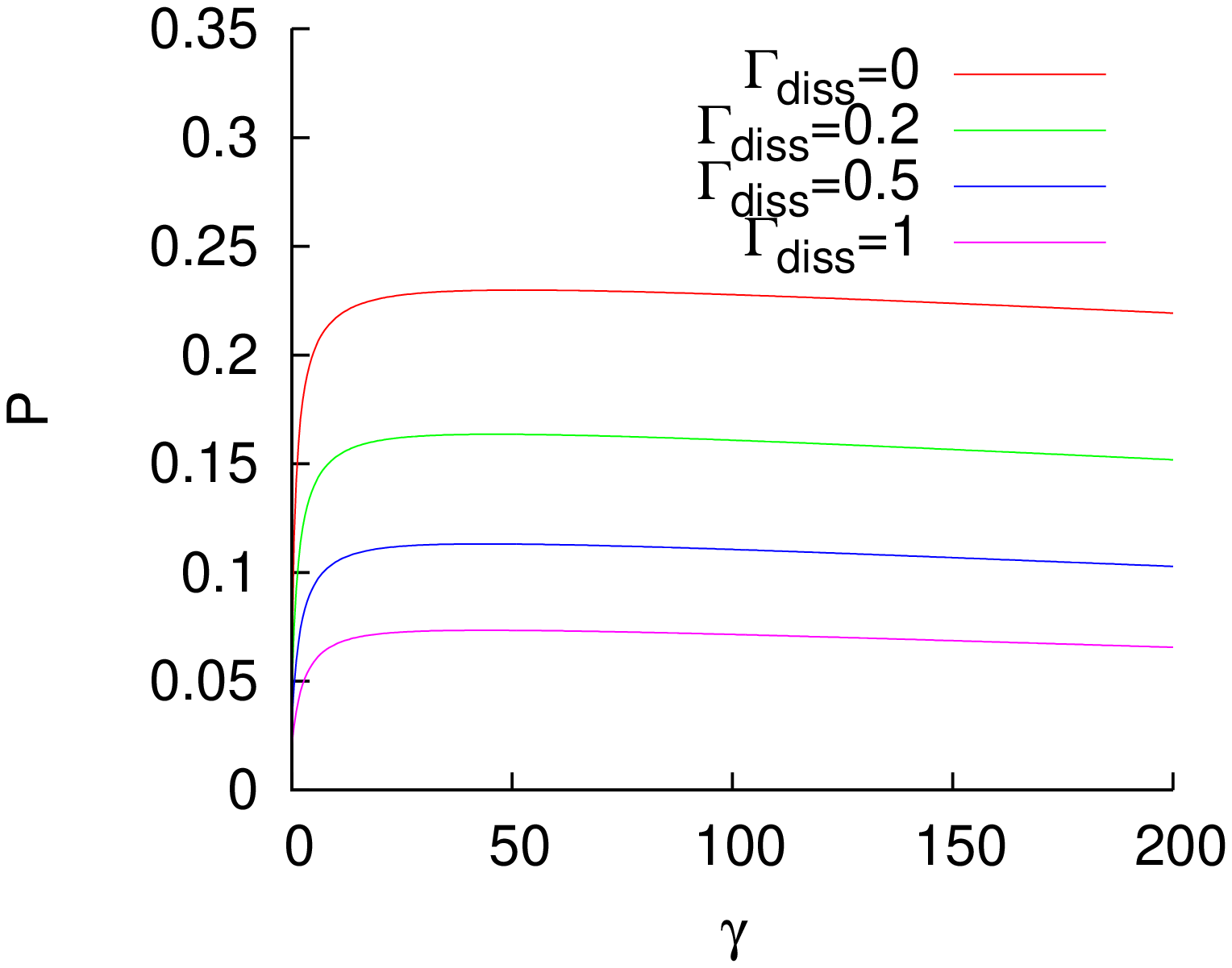} 
\includegraphics[scale=0.4]{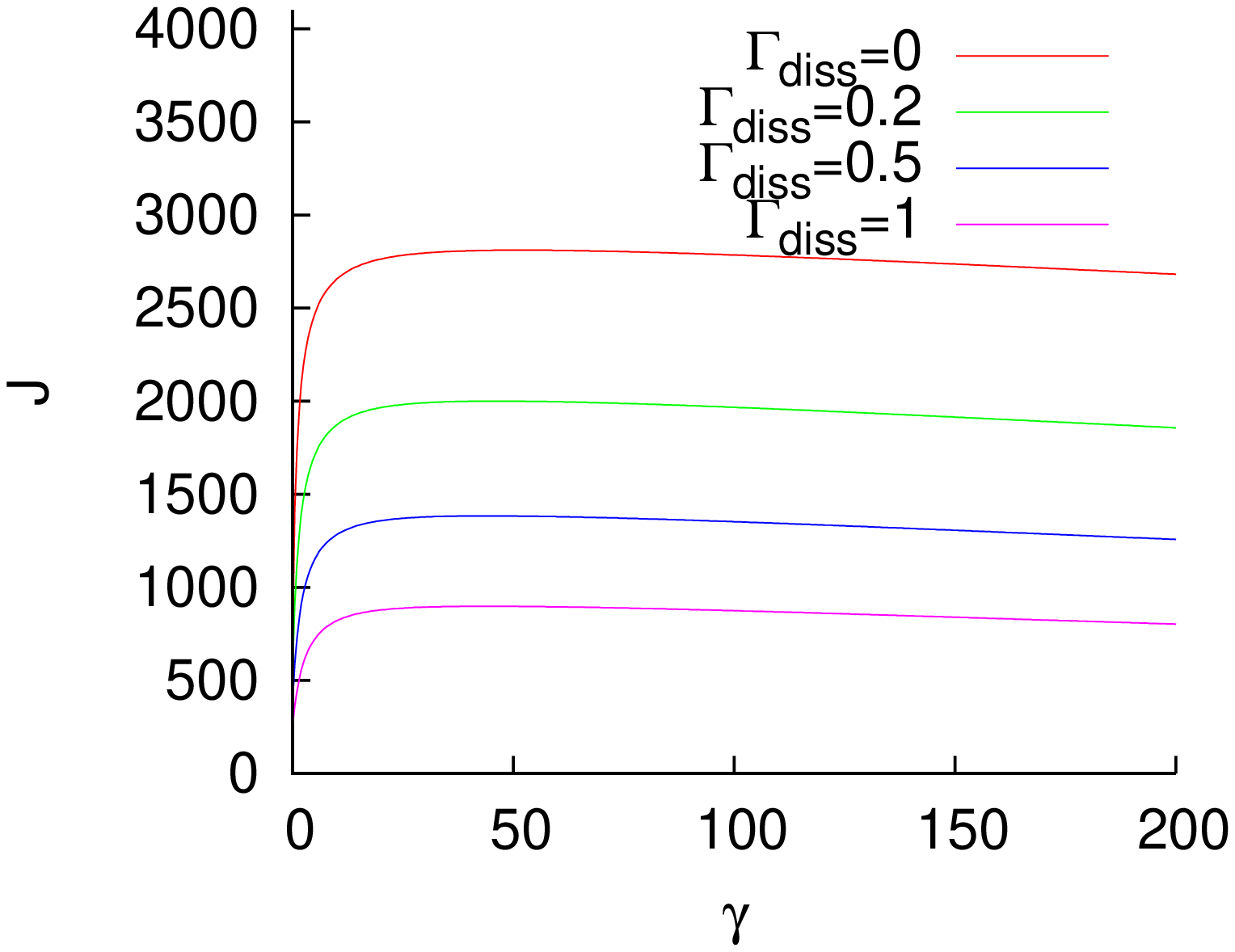} 
\ec
\caption{Energy (left) and excitation (right) fluxes  from the network to the sink for the FMO Hamiltonian 
as functions of the dephasing ratio,  $n=1$. Units of cm$^{-1}$.  }
\label{fig:fig2}
\end{figure}

In Fig. \ref{fig:fig2} the energy and excitation fluxes are plotted as a function of the dephasing ratio $\gamma$,  for 
$n=1$, and $\Gamma_{\text{diss}}=0,0.2,0.5,1\;\text{cm}^{-1}$.  The addition of  noise improves
both the excitation and the energy fluxes in this system and both measures of efficiency have practically the same  behaviour. Similar 
results arise for $n=100$, and are omitted. 
The optimal value of the dephasing ratio is equal for both the less and the highly excited scenarios, and for the 
energy and excitation transfers, with an optimal value $\gamma_{\text{FMO}}^*=60 \pm 1 \text{ cm}^{-1}$. It is also 
independent of the dissipation acting on the system. This result is of the same order, but quantitatively 
different, as the one obtained in \cite{wu:njp10} by using the mean trapping time as a measure of the efficiency and a 
global Redfield equation to describe the dynamics of the network. This optimal ratio is higher than in the case of the 
homogeneous network analysed before, 
due to the different order of magnitude of the ratio of the energies and the couplings. Again, the presence of dissipation in 
the system reduces the fluxes but it does not affect the qualitative behaviour. The similarity between both measures of efficiency 
comes from the differences between the one site energies and the couplings, which make the population of the third site the dominant 
component of Eq. (\ref{eq:flux}).

As the FMO Hamiltonian is inferred from experimental spectroscopy data, it is subject to experimental uncertainty. To check a more complete 
scenario, we performed a Monte Carlo simulation. For this simulation, 7000 random Hamiltonians were generated, where each parameter 
$x$ corresponds to a Gaussian distribution with mean in the corresponding FMO parameter $x_{\text{FMO}}$ and variance 
$\text{var}= 0.1\, x_{\text{FMO}}$.  By this simulation we analyse random Hamiltonians with the same order of magnitude as that of the FMO.

\begin{figure}[ht!]
\bc
\includegraphics[scale=0.4]{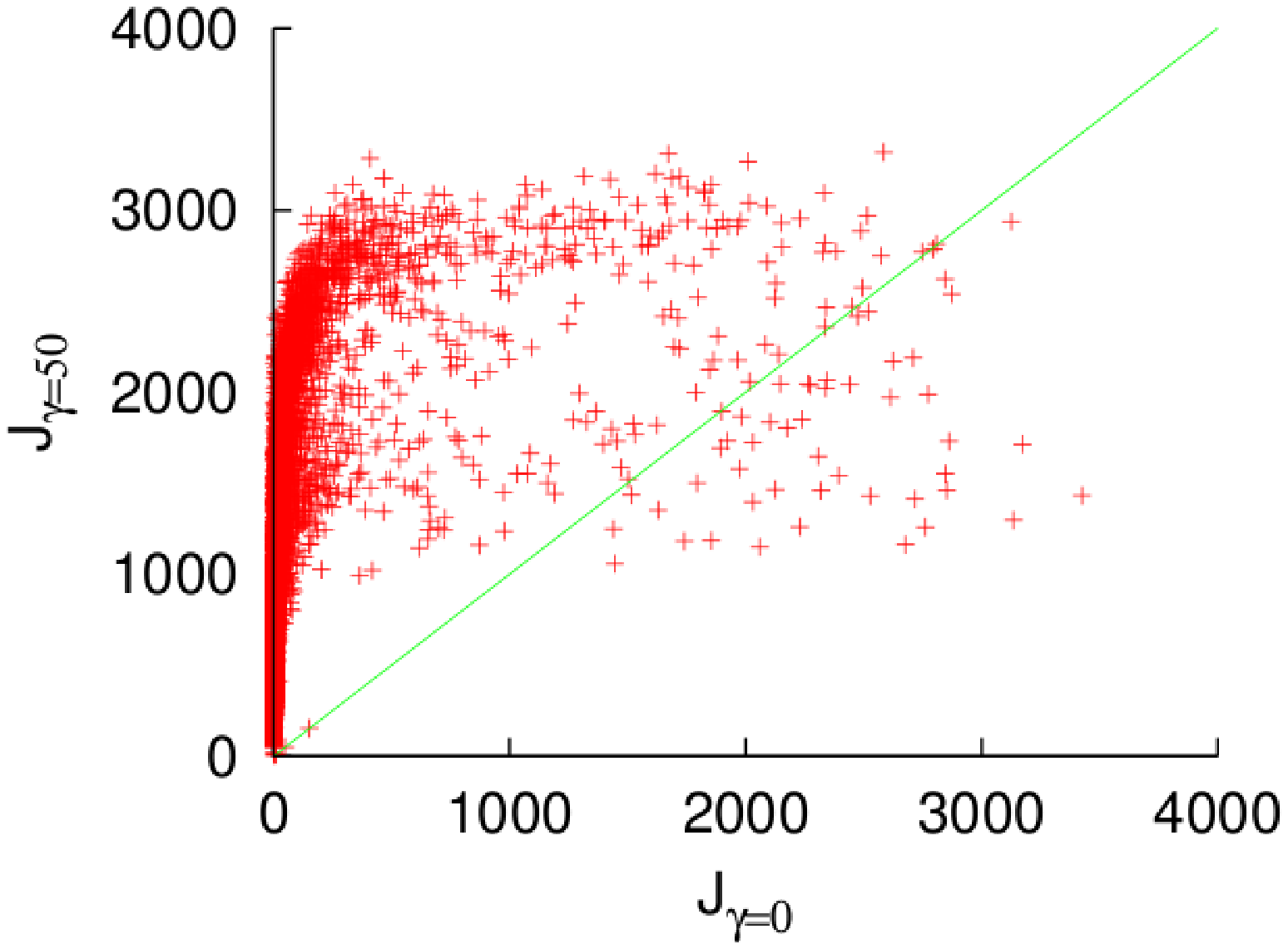} 
\includegraphics[scale=0.4]{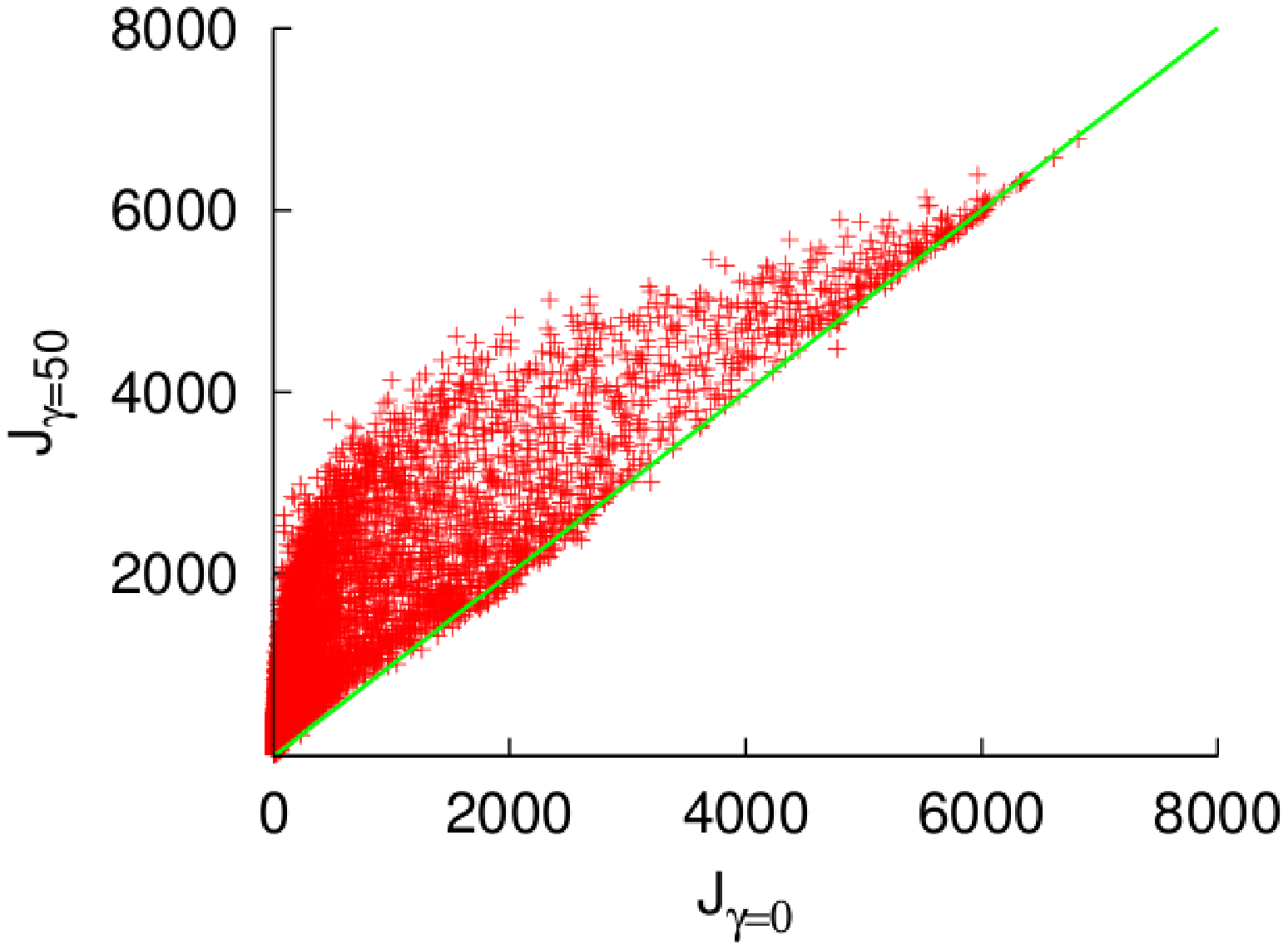} 
\ec
\caption{Energy flux for random Hamiltonians normally distributed around the FMO Hamiltonian, for $n=1$ (left)
and $n=100$ (right). The $y$-axis represents the energy flux under a dephasing channel with $\gamma=50$ and the $x$-axis represents the energy flux 
without dephasing. The green line represents the space where both fluxes are equal. Units of cm$^{-1}$.}
\label{fig:fig3}
\end{figure}

For a low temperature bath, the results for the energy transfer  are shown in Fig. \ref{fig:fig3}. For small dephasing, most of the 
configurations are improved. This improvement is more important for configurations with low efficiency, and it is less  for the most 
efficient ones. For higher values of  $\gamma$, most of the configurations are degraded, principally only the ones with low efficiencies are 
enhanced. Again, this result is similar if we use the population of the sink as our index. This similarity can be understood 
by analysing Eq. (\ref{eq:flux}). The energy flux depends on the population of the third site and on the coherences, modulated by the one site energy 
$\hbar \omega_3$ and the couplings $\hbar g_{3i}$, respectively.  If the one site energy of the third qubit is much higher than the couplings, the 
population of this site becomes the dominant term and both fluxes exhibit a similar behaviour. 

In this section, we proved that for the FMO complex, it is possible to improve both the energy and the excitation fluxes. That is due, 
principally, to the fact that the FMO Hamiltonian is a configuration with a very low efficiency. If the Hamiltonian is modified in order to obtain 
a similar but more efficient  one, this improvement reduces drastically. 

\section{Conclusions}
In this paper, we analysed the energy transfer  in  quantum networks and its 
behaviour when external noise is added to the system. Special emphasis has been placed on networks that model a real photosynthetic light-harvesting system, the FMO complex from green sulfur bacteria. This analysis has been performed in a steady state scenario. In this regime, the evolution of the system is not conditioned on the arrival of individual excitations, and the energy flows across the system continuously.

From our analysis we can conclude the following:

\begin{itemize}

\item Even in the non-equilibrium steady state, there are time independent coherences in the system. These coherences can 
contribute both in a positive or negative way to the energy transfer. 

\item The power of the system behaves in a very different way from the population transfer. That is due to 
the fact that the energy depends on the population but also on the coherences of the system. 

\item The population transfer is more amenable to improvement than the power. That means that even in the case in which the  
dephasing channel reduces the destructive interference in the system, it also reduces the coherences inside it, and that 
reduces the energy transfer in most cases. 

\item Also, the systems with low efficiency are more amenable to enhancement by a dephasing channel than the highly 
efficient ones. That is due to the fact that the interferences an be both destructive and constructive,  
and dephasing reduces both of them. Because of that, it can rarely improve a well performing configuration. 

\item Finally, systems that are in a high illumination regime are more stable under fluctuations due to the environmental 
interaction. 
\end{itemize}

From the biological point of view, and keeping in mind  that this is only a simple model far from the real photosynthetic scenario, 
we conclude that the energy transfer of the FMO complex can be enhanced by the addition of dynamical noise. On the other hand, this result is based 
on a single Hamiltonian inferred from empirical data. Our analysis shows also that if we slightly move from this Hamiltonian there 
are configurations that are better performing than the FMO. Also, we conclude that their energy transfer is not enhanced by a dephasing channel. 
So, in order to design a highly efficient light harvesting system, there are two possibilities: it can have low efficiency in a pure 
coherent dynamic and can be improved under the effect of decoherence, or it can be just more efficient for the coherent dynamics in the first place. The 
optimal choice depends on the environmental situation and on the practical constraints, but in most of the cases  analysed in 
this paper, the second choice is the optimal one.

\section{Acknowledgments}
The author would like  to thank M. Tiersch, G.G. Guerreschi, and H.J. Briegel for useful conversations.


\begin{thebibliography}{20}

\bibitem{engel:nature07}
Engel GS, Calhoun TR, Read EL, Ahn TK, Mancal T, Cheng YC, 
  Blankenship RE, Fleming GR (2007)
  Evidence for wavelike energy transfer through quantum coherence in
  photosyntetic systems.
   Nature 446: 782.

\bibitem{collini:n10}
Collini E, Wong CY, Wilk KE, Curmi PMG, Brumer P,   Scholes GD (2010)
  Coherently wired light-harvesting in photosynthetic marine algae at
  ambient temperature.
   Nature 463: 644--646.

\bibitem{panitchayangkoon:pnas10}
Panitchayangkoon G, Hayes D, Fransted KA, Caram JR, Harel E, Wen J,
  Blankenship RE,   Engel GS (2010)
  Long-lived quantum coherence in photosynthetic complexes at
  physiological temperature.
   Proc Natl Acad Sci 107: 12766.

\bibitem{plenio:njp08}
Plenio S Huelga M 2008
  Dephasing-assisted transport: quantum networks and biomolecules.
   J New Phys 10: 113019.

\bibitem{mohseni:jcp08}
Mohseni M, Rebentrost P, Lloyd S,   Aspuru-Guzik A (2008)
  Enviroment-assisted quantum walks in photosynthetic energy transfer.
   Journal of Chemical Physics 129: 174106.

\bibitem{olayacastro:prb08}
Olaya-Castro A, Lee CF, Olsen FF,   Johnson NF (2008)
  Efficiency of energy transfer in a light-harvesting system under
  quantum coherence.
   PRB 78: 085115.

\bibitem{caruso:jcp09}
Caruso F, Chin AW, Datta A, Huelga SF,   Plenio MB (2009)
  Highly efficient energy excitation transfer in light-harvesting
  complexes: The fundamental role of noise-assisted transport.
   J Chem Phys 131: 105106.

\bibitem{rebentrost:njp09}
Rebentrost P, Mohseni M, Kassal I, Lloyd S,   Aspuru-Guzik A (2009)
  Environment-assisted quantum transport.
   NJP 11: 033003.

\bibitem{chin:njp10}
Chin AW, Datta A, Caruso F, Huelga SF,   Plenio MB (2010)
  Noise-assisted energy transfer in quantum networks and light
  harvesting complexes.
   J New Phys 12: 065002.

\bibitem{wu:njp10}
Wu J, Liu F, Young Shen, Cao J,   Silbey RJ (2010)
  Efficient energy transfer in light-harvesting systems, I: Optimal
  temperature, reorganization energy and spatial-temporal correlations.
   NJP 12(105012).

\bibitem{scholak:pre11}
Scholak T, de Melo F, Wellens T, Mintert F,   Buchleitner A (2011)
  Efficient and coherent excitation transfer across disordered
  molecular networks.
   Phys Rev E 83(2): 021912.

\bibitem{scholak:jpb11}
Scholak T, Wellens T,   Buchleitner A (2011)
  Optimal networks for excitonic energy transport.
   J Phys B: At Mol Opt Phys 44: 184012.

\bibitem{mancal:njp10}
Mancal L, Valkunas T (2010)
  Exciton dynamics in photosynthetic complexes: excitation by coherent
  and incoherent light.
   NJP 12: 065044.

\bibitem{brumer:arxiv11}
Brumer M, Shapiro P (2012)
  Molecular response in one photon absorption: Coherent pulsed laser
  vs. thermal incoherent source.
 Proc Natl Acad Sci 109: 19575.

\bibitem{tiersch:ptrsa12}
Tiersch M, Popescu S,   Briegel HJ (2012)
  A critical view on transport and entanglement in models of
  photosynthesis.
   Phil Trans R Soc A 370: 3771.

\bibitem{manzano:arxiv11}
Manzano D, Tiersch M, Asadian A,   Briegel HJ (2011)
  Quantum transport efficiency and {F}ourier's law.
   arXiv: 1112.2839 [quant-ph].

\bibitem{asadian:arxiv12}
Asadian A, Manzano D, Tiersch M,   Briegel HJ (2012)
  Heat transport through lattices of quantum harmonic oscillators in
  arbitrary dimensions.
   arXiv: 1204.0904 [quant-ph].

\bibitem{blankenship:science11}
Blankenship RE, Tiede DM, Barber J, Brudvig GW, Fleming G, Ghirardi M,
  Gunner MR, Junge W, Kramer DM, Melis A, Moore TA, Moser, DG
  Nocera, Nozik AJ, Ort DR, Parson WW, Prince RC,   Sayre RT (2011)
  Comparing photosynthetic and photovoltaic efficiencies and
  recognizing the potential for improvement.
   Science 332(6031): 805--809.

\bibitem{caruso:pra10}
Caruso F, Chin AW, Datta A, Huelga SF,   Plenio MB (2010)
  Entanglement and entangling power of the dynamics in light-harvesting
  complexes.
   Phys Rev A 81(6): 062346.

\bibitem{breuer_02}
Breuer F Petruccione HP (2002)
   The theory of open quantum systems.
  Oxford: Oxford University Press. 

\bibitem{rivas:njp10}
Rivas A, Plato AD, Huelga S,   Plenio MB (2010)
  Markovian master equations: A critical study.
   J New Phys 12: 113032.

\bibitem{jacobs:pre12}
Jacobs K (2012)
  Quantum measurement and the first law of thermodynamics: The energy cost of measurement is the work value of the acquired information.
   Phys Rev E 86: 040106.


\bibitem{adolphs:bj06}
Adolphs T, Renger J (2006)  
  How proteins trigger excitation energy transfer in the {FMO} complex
  of green sulfur bacteria.
   Biophys Journal 91(2778).

\bibitem{cao:jpc97}
Cao JS (1997)  
  A phase-space study of Bloch--Redfield theory.
   Journal of Chemical Physics 107: 3204.

\bibitem{linden:prl10}
Linden N, Popescu S,   Skrzypczyk P (2010) 
  How small can thermal machines be? {T}he smallest possible
  refrigerator.
   Phys Rev Lett 105(13): 130401.

\end{thebibliography}
\end{document}